\documentclass[11pt,a4paper]{article}

\usepackage{cite}

\usepackage{amsmath}

\usepackage[dvipdfmx]{graphicx}

\usepackage{subfig}

\usepackage{array}

\usepackage{amsfonts}

\usepackage{epsf}

\usepackage{epsfig}

\usepackage{amssymb}

\usepackage{bbm}

\usepackage{delarray}

\usepackage{caption}

\usepackage{amstext}

\usepackage{graphics}

\usepackage{footnote}

\usepackage[T1]{fontenc}
\usepackage[bitstream-charter]{mathdesign}

\usepackage[latin1]{inputenc}

\usepackage{xcolor}
\definecolor{bl}{rgb}{0.0,0.2,0.6}

\usepackage{graphicx}
\usepackage{sectsty}
\usepackage[compact]{titlesec}
\allsectionsfont{\color{bl}\scshape\selectfont}

\makeatletter
\def\printtitle{%
    {\color{bl} \centering \huge \sc \textbf{\@title}\par}}
\makeatother

\title{\\ \large \vspace*{-10pt} PT-/non-PT-Symmetric and non-Hermitian Hellmann Potential: Approximate Bound and Scattering
States with Any $\ell$-Values\vspace*{10pt}}

\makeatletter
\def\printauthor{%
    {\centering \small \@author}}
\makeatother

\author{%
    Altuð Arda \\
    arda@hacettepe.edu.tr \\
    Ramazan Sever \\
    sever@metu.edu.tr \\
    \vspace{20pt}
    }

\usepackage{fancyhdr}
\usepackage{lastpage}
\usepackage[runin]{abstract}
\abslabeldelim{\quad}
\catcode`ð=\active
 \defð{\u{g}}
 \catcode`Ð=\active
 \defÐ{\u{G}}
 \catcode`Ý=\active
 \defÝ{\. I}
 \catcode`ö=\active
 \defö{\"{o}}
 \catcode`Ö=\active
 \defÖ{\"O}
 \catcode`ü=\active
 \defü{\"{u}}
 \catcode`Ü=\active
 \defÜ{\"{U}}
 \catcode`Þ=\active
 \defÞ{\c{S}}
 \catcode`þ=\active
 \defþ{\c{s}}
 \catcode`ý=\active
 \defý{{\i}}
 \catcode`ç=\active
 \defç{\d{c}}
 \catcode`Ç=\active
 \defÇ{\d{C}}

\makesavenoteenv{tabular}
\begin{document}

\printtitle

\printauthor

\begin{abstract}
We investigate the approximate bound state solutions of the
Schrödinger equation for the PT-/non-PT-symmetric and non
Hermitian Hellmann potential. Exact energy eigenvalues and
corresponding normalized wave functions are obtained. Numerical
values of  energy eigenvalues for the bound states are compared
with the ones obtained before. Scattering state solutions are also
studied. Phase shifts of the potential are written in terms of the
angular momentum quantum number $\ell$.
\end{abstract}

\section{Introduction}
PT-symmetric quantum mechanics has been widely studied in recent
years. The usual form of quantum mechanics has the Hamiltonian
defining symmetries of the system is Hermitian. However for
PT-symmetric case, it has real spectra although it is not
Hermitian. Bender and Boettcher \cite{b1} studied this case. Later
many authors studied PT-symmetric and non-Hermitian cases having
real and/or complex eigenvalues
\cite{b2,b3,jzs,km,b4,jll,lcv,bcq,b5}.

The Hellmann potential
\begin{eqnarray}
V(r)=\frac{1}{r}\left(-a+be^{-\lambda r}\right)\,,
\end{eqnarray}
with $b>0$ was first proposed by Hellmann \cite{h1,h2} (then
called as the 'Hellmann potential' independently of the sign of
$b$) which has many applications in atomic physics and condensed
physics \cite{h3}. The Hellmann potential has been used as a
potential model to calculate the electronic wave functions of
metals and semiconductors \cite{jl}. Many authors have studied the
electron-core \cite{c1,i1,cl} and electron-ion \cite{lcv}
interactions by using this potential. In Ref. \cite{dc}, it has
been proposed that the Hellmann potential is a suitable ground for
study of inner-shell ionization problems. The present potential
could be used as a potential model for the alkali hydride
molecules \cite{vs}.

Energy eigenvalues of the Hellmann potential have recently been
studied by various authors with the help of different methods such
as $1/N$ expansion method \cite{st}, shifted large-$N$ expansion
method \cite{dmy}, the method of potential envelopes \cite{bdy},
the $J$-matrix approach [24] and the generalized Nikiforov-Uvarov
method \cite{hta, ha}. In the present work, we solve the
Shrödinger-Hellmann problem in terms of the hypergeometric
functions by using an approximation instead of the centrifugal
term. We extend also the computation including the solutions of
the Hellmann-like potential having the form
\begin{eqnarray}
V(x)=-\frac{a}{x}+\frac{b}{x}\,e^{-\lambda x}\,,
\end{eqnarray}
which can be written in a PT-symmetric form and the energy spectra
and eigenfunctions of PT-/non-PT- and non-Hermitian Hellmann
potential are obtained with any angular momentum. Scattering state
solutions are also studied. Phase shifts of the potential are
written in terms of the angular momentum quantum number $\ell$.

The organization of this work is as follows. In Section 2, we find
the approximate energy eigenvalues and the corresponding
normalized wave functions of the Hellmann potential. In Section 3,
we obtain the phase shifts of the potential under consideration in
terms of the quantum number $\ell$. We give our conclusions in
last section.

\section{Bound States}
\textit{i. Radial Solutions}

The radial part of the Schrödinger equation (SE) \cite{s1}
\begin{eqnarray}
\left\{\frac{d^2}{dr^2}-\frac{\ell(\ell+1)}{r^2}+\frac{2m}{\hbar^2}\left[E-V(r)\right]\right\}\mathcal{R}(r)=0\,.
\end{eqnarray}
where $\mathcal{R}(r)=R(r)/r$, $\ell$ is the angular momentum
quantum number, $m$ is the particle mass moving in the potential
field $V(r)$ and $E$ is the nonrelativistic energy of particle.

Using the following approximation instead of the centrifugal term
\cite{hc}
\begin{eqnarray}
\frac{1}{r^2}\sim \frac{\lambda^2}{(1-e^{-\lambda r})^2}\,,
\end{eqnarray}
inserting Eq. (1) into Eq. (3) and defining a new variable
$u=e^{-\lambda r}$ ($0\leq u \leq 1$), Eq. (3) becomes
\begin{eqnarray}
&&u(1-u)\frac{d^2R(u)}{du^2}+(1-u)\frac{dR(u)}{du}\nonumber\\&\times&
\left[-\frac{\ell(\ell+1)}{1-u}+\left(\frac{2mE}{\lambda^2\hbar^2}
+\frac{2ma\lambda}{\lambda^2\hbar^2}-\ell(\ell+1)\right)\,\frac{1}{u}-
\frac{2mb\lambda}{\lambda^2\hbar^2}-\frac{2mE}{\lambda^2\hbar^2}\right]R(u)=0\,.\nonumber\\
\end{eqnarray}
Taking a trial wave function as
\begin{eqnarray}
R(u)=u^{\lambda_{1}}(1-u)^{\lambda_{2}}\psi(u)\,,
\end{eqnarray}
and inserting into Eq. (5), we obtain
\begin{eqnarray}
&&u(1-u)\frac{d^2\psi(u)}{du^2}+\left[1+2\lambda_{1}-(2\lambda_{1}
+2\lambda_{2}+1)u\right]\frac{d\psi(u)}{du}\nonumber\\
&\times&\left[-\lambda^2_{1}-\lambda^2_{2}-2\lambda_{1}\lambda_{2}-\frac{2m}
{\lambda^2\hbar^2}(E+b\lambda)\right]\psi(u)=0\,,
\end{eqnarray}
where
\begin{eqnarray}
\lambda^2_{1}&=&-\frac{2m}{\lambda^2\hbar^2}(E+a\lambda)+\ell(\ell+1)\,,\\
\lambda_{2}&=&\frac{1}{2}\left[1 \pm
\sqrt{1+4\ell(\ell+1)\,}\,\right]\,.
\end{eqnarray}
Comparing Eq. (7) with the hypergeometric equation of the
following form  \cite{as}
\begin{eqnarray}
u(1-u)\frac{d^2y(u)}{du^2}+[c'-(a'+b'+1)u]\frac{dy(u)}{du}-a'b'y(u)=0\,,
\end{eqnarray}
we find the solution of Eq. (7) as the hypergeometric function
\begin{eqnarray}
\psi(u)=\,_2F_1(a',b';c';u)\,,
\end{eqnarray}
with
\begin{eqnarray}
a'&=&\lambda_{1}+\lambda_{2}+\Lambda_{1}\,,\\
b'&=&\lambda_{1}+\lambda_{2}-\Lambda_{1}\,,\\
c'&=&1+2\lambda_{1}\,.
\end{eqnarray}
where
$\Lambda_{1}=\frac{1}{2}\sqrt{-\frac{8m}{\lambda^2\hbar^2}(E+b\lambda)\,}$.

The total wave functions in Eq. (6) are given as
\begin{eqnarray}
R(u)=\mathcal{N}u^{\lambda_{1}}(1-u)^{\lambda_{2}}\,_2F_1(a',b';c';u)\,.
\end{eqnarray}
where $\mathcal{N}$ is normalization constant and will be
calculated below. When either $a'$ or $b'$ equals to a negative
integer $-n$, the hypergeometric function $\psi(u)$ can give a
finite solution form. This gives us a polynomial of degree $n$ in
Eq. (11) and from the following quantum condition
\begin{eqnarray}
-n=\lambda_{1}+\lambda_{2}+\frac{1}{2}\sqrt{-\frac{8m}{\lambda^2\hbar^2}(E+b\lambda)\,}\,,\,\,\,(n=0,
1, 2, \ldots)
\end{eqnarray}
the energy eigenvalue becomes
\begin{eqnarray}
E&=&-\frac{1}{8m\hbar^2(n+\ell+1)^2}\bigg\{4m^2(a^2+b^2)+4m\hbar^2\lambda
b\left[2\ell^2+(n+\ell)^2+\ell(3+2n)\right]\nonumber\\&+&\lambda^2\hbar^4\left[\ell(1+2n)+(n+\ell)^2\right]^2
+4am\left[-2bm+\lambda\hbar^2\left[\ell(1+2n)+(n+\ell)^2\right]\right]\bigg\}\,.\nonumber\\
\end{eqnarray}
The numerical results obtained from last equation are listed in
Table I. We also compare them with the ones given in two different
papers. They are in agreement with those of the previous results
where we should also stress that our results are consistent with
the ones given in Ref. \cite{hta}. Eq. (16) gives the wave
functions as
\begin{eqnarray}
R(u)=\mathcal{N}u^{\lambda_{1}}(1-u)^{\lambda_{2}}\,_{2}F_{1}(-n,n+2\lambda_{1}+2\lambda_{2};1+2\lambda_{1};u)\,,
\end{eqnarray}
where the normalization constant is obtained from
$\int_{0}^{1}|R(u)|^2du=1$ which can be written as
\begin{eqnarray}
&&|\mathcal{N}|^2\frac{\Gamma(1+2\lambda_{1})}{m!\Gamma(-n)\Gamma(n+2\lambda_{1}+2\lambda_{2})}
\sum_{m=0}^{\infty}\frac{\Gamma(-n+m)\Gamma(n+2\lambda_{1}+2\lambda_{2}+m)}{\Gamma(1+2\lambda_{1}+m)}\nonumber\\&\times&
\int_{0}^{1}u^{m+2\lambda_{1}}(1-u)^{2\lambda_{2}}\,_{2}F_{1}(-n,n+2\lambda_{1}+2\lambda_{2};1+2\lambda_{1};u)du=1\,.
\end{eqnarray}
We use the following representation of the hypergeometric
functions \cite{as}
\begin{eqnarray}
\,_{2}F_{1}(p,q;r;z)=\frac{\Gamma(r)}{\Gamma(p)\Gamma(q)}\sum_{m=0}^{\infty}
\frac{\Gamma(p+m)\Gamma(q+m)}{\Gamma(r+m)}\frac{z^{m}}{m!}\,,
\end{eqnarray}
By using the following identity \cite{gm}
\begin{eqnarray}
\int_{0}^{1}s^{\nu-1}(1-s)^{\mu-1}\,_{2}F_{1}(\alpha,\beta;\gamma;as)ds
=\frac{\Gamma(\mu)\Gamma(\nu)}{\Gamma(\mu+\nu)}\,_{3}F_{2}(\nu,\alpha,\beta;\mu+\nu;\gamma;a)\,,
\end{eqnarray}
we obtain the normalization constant as
\begin{eqnarray}
|\mathcal{N}|^2&=&\frac{\Gamma(1+2\lambda_{1})\Gamma(1+2\lambda_{2})}{m!\Gamma(-n)\Gamma(n+2\lambda_{1}+2\lambda_{2})}
\sum_{m=0}^{\infty}\frac{\Gamma(-n+m)\Gamma(n+2\lambda_{1}+2\lambda_{2}+m)}{\Gamma(1+2\lambda_{1}+m)}
\nonumber\\&\times&\,_{3}F_{2}(1+2\lambda_{1}+m,-n,n+2\lambda_{1}
+2\lambda_{2};2+2\lambda_{1}+2\lambda_{2}+m;1+2\lambda_{1};1)\,.\nonumber\\
\end{eqnarray}

For the completeness, we give the energy eigenvalues of the
Coulomb potential which corresponds to the case where $b=0$. We
write the energy levels of this potential from Eq. (17) as
($\lambda=0$, $\hbar=1$)
\begin{eqnarray}
E=-\frac{ma^2}{2(n+\ell+1)^2}\,.
\end{eqnarray}
\textit{ii. PT-Symmetric Solutions}

Inserting Eq. (2) into the following one-dimensional SE \cite{s1}
\begin{eqnarray}
-\frac{\hbar^2}{2m}\,\frac{d^2\phi(x)}{dx^2}+\left[V(x)-E\right]\phi(x)=0\,.
\end{eqnarray}
using the following approximation instead of $1/x$ in the
potential (see, Fig. 1)
\begin{eqnarray}
\frac{1}{x}\sim \frac{\lambda}{1-e^{-\lambda x}}\,,
\end{eqnarray}
and taking a new variable as $u=1/(1-e^{-\lambda x})$ we obtain
\begin{eqnarray}
&&u(1-u)\frac{d^2\phi(u)}{du^2}+(1-2u)\frac{d\phi(u)}{du}\nonumber\\&
\times&\left[\left(\frac{2ma}{\lambda\hbar^2}+\frac{2mE}{\lambda^2\hbar^2}\right)\frac{1}{1-u}
+\left(\frac{2mb}{\lambda\hbar^2}+\frac{2mE}{\lambda^2\hbar^2}\right)\frac{1}{u}\right]\phi(u)=0\,.
\end{eqnarray}
Defining the wave function as
\begin{eqnarray}
\phi(u)=u^{\lambda_{1}}(1-u)^{\lambda_{2}}\psi(u)\,,
\end{eqnarray}
and following the same procedure in the above we get
\begin{eqnarray}
&&u(1-u)\frac{d^2\psi(u)}{du^2}+\left[1+2\lambda_{1}-(2\lambda_{1}+2\lambda_{2}+1)u\right]\frac{d\psi(u)}{du}\nonumber\\
&-&\left[\lambda_{1}(\lambda_{1}+1)+\lambda_{2}(\lambda_{2}+1)+2\lambda_{1}\lambda_{2}\right]\psi(u)=0\,,
\end{eqnarray}
where
\begin{eqnarray}
\lambda^2_{1}&=&-\frac{2m}{\lambda\hbar^2}\,(b+\frac{E}{\lambda})\,,\\
\lambda^2_{2}&=&-\frac{2m}{\lambda\hbar^2}\,(a+\frac{E}{\lambda})\,.
\end{eqnarray}
The solution of Eq. (28) and the total function is given as,
respectively,
\begin{eqnarray}
&\psi(u)\sim\,_2F_1(a',b';c';u)\,,\\
&\phi(u)\sim
u^{\lambda_{1}}(1-u)^{\lambda_{2}}\,_2F_1(a',b';c';u)\,,
\end{eqnarray}
where
\begin{eqnarray}
a'&=&1+\lambda_{1}+\lambda_{2}\,,\\
b'&=&\lambda_{1}+\lambda_{2}\,,\\
c'&=&1+2\lambda_{1}\,.
\end{eqnarray}
The energy eigenvalues are written as
\begin{eqnarray}
E=-\frac{\lambda}{4A(1+n)^2}\bigg\{a^2A^2+2aA[-Ab+(1+n)^2]+[Ab+(1+n)^2]^2\bigg\}\,.
\end{eqnarray}
where $A=2m/\lambda\hbar^2$.

\textit{iii. Non-Hermitian PT-Symmetric Form}

Changing the potential parameters as $a\rightarrow ia,
b\rightarrow ib, \beta\rightarrow i\beta$ in Eq. (2), the
potential satisfies
\begin{eqnarray}
V^{*}(-x)=\left(\frac{ia}{x}-\frac{ib}{x}\,e^{i\beta
x}\right)^{*}=V(x)\,,
\end{eqnarray}
which shows that we obtain non-Hermitian PT-symmetric form of the
Hellmann-like potential. Inserting Eq. (37) into Eq. (24) and
using the variable $u=[1-e^{-i\lambda x}]^{-1}$, we obtain
\begin{eqnarray}
&&u(1-u)\frac{d^2\phi(u)}{du^2}+(1-2u)\frac{d\phi(u)}{du}\nonumber\\&
\times&\left[\left(\frac{2ma}{\lambda\hbar^2}-\frac{2mE}{\lambda^2\hbar^2}\right)\frac{1}{1-u}
+\left(\frac{2mb}{\lambda\hbar^2}-\frac{2mE}{\lambda^2\hbar^2}\right)\frac{1}{u}\right]\phi(u)=0\,.
\end{eqnarray}
where
\begin{eqnarray}
\lambda^2_{1}&=&-\frac{2m}{\lambda\hbar^2}\,(b-\frac{E}{\lambda})\,,\\
\lambda^2_{2}&=&-\frac{2m}{\lambda\hbar^2}\,(a-\frac{E}{\lambda})\,.
\end{eqnarray}
Following the same steps, we obtain the wave function for the
non-Hermitian PT-symmetric Hellmann-like potential
\begin{eqnarray}
\phi(u)\sim
u^{\lambda_{1}}(1-u)^{\lambda_{2}}\,_2F_1(a',b';c';u)\,,
\end{eqnarray}
and the energy spectrum as
\begin{eqnarray}
E&=&-\frac{1}{8m\hbar^2(1+n)^2}\bigg\{4m^2a^2\nonumber\\&+&4ma[-2mb+\lambda
\hbar^2(1+n)^2]+[2mb+\lambda\hbar^2(1+n)^2]^2\bigg\}\,.\nonumber\\
\end{eqnarray}
\textit{iv. Non-Hermitian Non-PT-Symmetric Form}

Case 1: $a$ and $b$ real, $\lambda \rightarrow i\lambda$

In this case the potential satisfies $[V(x)]^{*}\neq V(x)$ so it
has a non-Hermitian non-PT-symmetric form given as
\begin{eqnarray}
V(x)=-ia\lambda\,\frac{1}{1-e^{-i\lambda
x}}+ib\lambda\,\frac{e^{-i\lambda x}}{1-e^{-i\lambda x}}\,,
\end{eqnarray}
Using the variable $u=[1-e^{-i\lambda x}]^{-1}$, we obtain
\begin{eqnarray}
&&u(1-u)\frac{d^2\phi(u)}{du^2}+(1-2u)\frac{d\phi(u)}{du}\nonumber\\&
\times&\left[-\left(\frac{2mia}{\lambda\hbar^2}+\frac{2mE}{\lambda^2\hbar^2}\right)\frac{1}{1-u}
-\left(\frac{2mb}{\lambda\hbar^2}+\frac{2mE}{\lambda^2\hbar^2}\right)\frac{1}{u}\right]\phi(u)=0\,.
\end{eqnarray}
with
\begin{eqnarray}
\lambda^2_{1}&=&\frac{2m}{\lambda\hbar^2}\,(ib+\frac{E}{\lambda})\,,\\
\lambda^2_{2}&=&\frac{2m}{\lambda\hbar^2}\,(ia+\frac{E}{\lambda})\,.
\end{eqnarray}
Following the same steps, we obtain the wave function for the
non-Hermitian non-PT-symmetric Hellmann-like potential
\begin{eqnarray}
\phi(u)\sim
u^{\lambda_{1}}(1-u)^{\lambda_{2}}\,_2F_1(a',b';c';u)\,,
\end{eqnarray}
and the energy spectrum as
\begin{eqnarray}
E&=&-\frac{1}{8m\hbar^2(1+n)^2}\bigg\{4m^2a^2\nonumber\\&-&4ma[2mb-i
\lambda\hbar^2(1+n)^2]+[2mb+i\lambda\hbar^2(1+n)^2]^2\bigg\}\,.\nonumber\\
\end{eqnarray}

Case 2: $\beta$ real, $a \rightarrow ia$ and $b \rightarrow ib$

By using the variable $t=1/(1-e^{-\beta x})$ we obtain the
following in the present case
\begin{eqnarray}
V(x)=-ia\lambda\,\frac{1}{1-e^{-\lambda
x}}+ib\lambda\,\frac{e^{-\lambda x}}{1-e^{-\lambda x}}\,,
\end{eqnarray}
Using the variable $u=[1-e^{-\lambda x}]^{-1}$, we obtain
\begin{eqnarray}
&&u(1-u)\frac{d^2\phi(u)}{du^2}+(1-2u)\frac{d\phi(u)}{du}\nonumber\\&
\times&\left[\left(\frac{2mia}{\lambda\hbar^2}+\frac{2mE}{\lambda^2\hbar^2}\right)\frac{1}{1-u}
+\left(\frac{2mib}{\lambda\hbar^2}+\frac{2mE}{\lambda^2\hbar^2}\right)\frac{1}{u}\right]\phi(u)=0\,.
\end{eqnarray}
where
\begin{eqnarray}
\lambda^2_{1}&=&-\frac{2m}{\lambda\hbar^2}\,(ib+\frac{E}{\lambda})\,,\\
\lambda^2_{2}&=&-\frac{2m}{\lambda\hbar^2}\,(ia+\frac{E}{\lambda})\,.
\end{eqnarray}
Following the same steps as in the above section we obtain the
wave function as
\begin{eqnarray}
\phi(u)\sim
u^{\lambda_{1}}(1-u)^{\lambda_{2}}\,_2F_1(a',b';c';u)\,,
\end{eqnarray}
and the energy spectrum as
\begin{eqnarray}
E&=&\frac{1}{8m\hbar^2(1+n)^2}\bigg\{4m^2a^2\nonumber\\&-&4ma[2mb+i\lambda
\hbar^2(1+n)^2]+[2mb-i\lambda\hbar^2(1+n)^2]^2\bigg\}\,.\nonumber\\
\end{eqnarray}
We now study the scattering state solutions of the Hellmann
potential in the next section.

\section{Scattering States and Phase Shifts}
In order to obtain the scattering state solutions we choose the
variable as $t=1-e^{-\lambda r}$ in Eq. (3) and we get the
following equation in terms of $t$
\begin{eqnarray}
&&t(1-t)\frac{d^2R(t)}{dt^2}-t\frac{dR(t)}{dt}\nonumber\\
&+&\left[[\frac{2m}{\lambda\hbar^2}\,(a-\epsilon)-\ell(\ell+1)]
\frac{1}{1-t}-\ell(\ell+1)\,\frac{1}{t}-\frac{2m}{\lambda\hbar^2}\,(b-\epsilon)\right]R(t)=0\,,\nonumber\\
\end{eqnarray}
where $-\epsilon=E/\lambda$. By using a trial wave function as
$R(t)=t^{\mu}(1-t)^{\nu}\psi(t)$, we get
\begin{eqnarray}
&&t(1-t)\frac{d^2\psi(t)}{dt^2}\nonumber\\&+&[2\mu-(2\mu+2\nu+1)t]\frac{d\psi(t)}{dt}
+\left[\mu^2+\nu^2+2\mu\nu+\frac{2m}{\lambda\hbar^2}\,(b-\epsilon)\right]\psi(t)=0\,,\nonumber\\
\end{eqnarray}
with
\begin{eqnarray}
\mu=\left\{
\begin{array}{l}
-\ell \\
1+\ell\\
\end{array}\right.
\,;\,\,\,\nu=-i\kappa\,;\,\,\,\kappa=\sqrt{\frac{2m}{\lambda\hbar^2}\,(a-\epsilon)-\ell(\ell+1)\,}\,.
\end{eqnarray}

By using the following abbreviations
\begin{eqnarray}
\xi_{1}&=&\mu-i\kappa+\Lambda_{2}\,,\\
\xi_{2}&=&\mu-i\kappa-\Lambda_{2}\,,\\
\xi_{3}&=&2\mu\,,
\end{eqnarray}
where
$\Lambda_{2}=\sqrt{\frac{2m}{\lambda\hbar^2}\,(\epsilon-b)\,}$,
Eq. (56) can be written as a hypergeometric-type equation
\cite{gm}
\begin{eqnarray}
t(1-t)\frac{d^2\psi(t)}{dt^2}+[\xi_{3}-(\xi_{1}+\xi_{2}+1)t]\frac{d\psi(t)}{dt}-\xi_{1}\xi_{2}\psi(t)=0\,.
\end{eqnarray}
Its solution is given by
\begin{eqnarray}
\psi(t)=\,_{2}F_{1}(\xi_{1},\xi_{2};\xi_{3};t)\,,
\end{eqnarray}
where the parameters satisfy the followings
\begin{eqnarray}
\xi_{3}-\xi_{1}-\xi_{2}=(\xi_{1}+\xi_{2}-\xi_{3})^{*}\,;\,\,\xi_{3}-\xi_{1}=\xi^{*}_{2}\,;\,\,\xi_{3}-\xi_{2}=\xi^{*}_{1}\,.
\end{eqnarray}
which are used in determination of the phase shifts. By using the
following equality of the hypergeometric functions \cite{gm}
\begin{eqnarray}
&&_{2}F_{1}(a'',b'';c'';z)=\frac{\Gamma(c'')\Gamma(c''-a''-b'')}
{\Gamma(c''-a'')\Gamma(c''-b'')}\,_{2}F_{1}(a'',b'';a''+b''-c''+1;1-z)\nonumber\\
&+&(1-z)^{c''-a''-b''}\times\nonumber\\ &&\frac{\Gamma(c'')\Gamma(a''+b''-c'')}{\Gamma(a'')\Gamma(b'')}\,_{2}F_{1}
(c''-a'',c''-b'';c''-a''-b''+1;1-z)\,,\nonumber\\
\end{eqnarray}
and $\,_{2}F_{1}(a'',b'';c'';0)=1$, we write the solution of Eq.
(61) in the limit of $r \rightarrow \infty$ as
\begin{eqnarray}
&&_{2}F_{1}(\xi_{1},\xi_{2};\xi_{3};1-e^{-\lambda r})
\xrightarrow[r \to \infty]{}\nonumber\\ &&
\frac{\Gamma(2\mu)\Gamma(2i\kappa)}{\Gamma\left(\mu+i\kappa-
\sqrt{\frac{2m}{\lambda\hbar^2}\,(\epsilon-b)\,}\,\right)\Gamma\left(\mu+i\kappa+
\sqrt{\frac{2m}{\lambda\hbar^2}\,(\epsilon-b)\,}\,\right)}\nonumber\\&+&e^{-2i\kappa\lambda
r}\frac{\Gamma(2\mu)\Gamma(-2i\kappa)}{\Gamma\left(\mu-i\kappa+
\sqrt{\frac{2m}{\lambda\hbar^2}\,(\epsilon-b)\,}\,\right)\Gamma\left(\mu-i\kappa-
\sqrt{\frac{2m}{\lambda\hbar^2}\,(\epsilon-b)\,}\,\right)}\,.
\end{eqnarray}
Defining the following
\begin{eqnarray}
\frac{\Gamma(c''-a''-b'')}{\Gamma(c''-a'')\Gamma(c''-b'')}=
\left|\frac{\Gamma(c''-a''-b'')}{\Gamma(c''-a'')\Gamma(c''-b'')}\right|e^{i\delta}\,,
\end{eqnarray}
and also with the help of Eq. (63)
\begin{eqnarray}
\left(\frac{\Gamma(c''-a''-b'')}{\Gamma(c''-a'')\Gamma(c''-b'')}\right)^{*}
=\left|\frac{\Gamma(c''-a''-b'')}{\Gamma(c''-a'')\Gamma(c''-b'')}\right|e^{-i\delta}\,,
\end{eqnarray}
Eq. (65) becomes
\begin{eqnarray}
&&_{2}F_{1}(\xi_{1},\xi_{2};\xi_{3};1-e^{-\lambda r}) \xrightarrow[r
\to \infty]{}\nonumber\\
&&\Gamma(2\mu)\left|\frac{\Gamma(2i\kappa)}{\Gamma\left(\mu+i\kappa-
\sqrt{\frac{2m}{\lambda\hbar^2}\,(\epsilon-b)\,}\,\right)\Gamma\left(\mu+i\kappa+
\sqrt{\frac{2m}{\lambda\hbar^2}\,(\epsilon-b)\,}\,\right)}\right|\nonumber\\&\times&
e^{-i\kappa\lambda r}\left[e^{i(\delta+\kappa\lambda
r)}+e^{-i(\delta+\kappa\lambda r)}\right]\,.
\end{eqnarray}
We write the total wave function wi the help of this result as
\begin{eqnarray}
R(r \rightarrow
\infty)&=&2\Gamma(2\mu)\left|\frac{\Gamma(2i\kappa)}{\Gamma\left(\mu+i\kappa-
\sqrt{\frac{2m}{\lambda\hbar^2}\,(\epsilon-b)\,}\,\right)\Gamma\left(\mu+i\kappa+
\sqrt{\frac{2m}{\lambda\hbar^2}\,(\epsilon-b)\,}\,\right)}\right|\nonumber\\&\times&sin\left(\delta+\lambda\kappa
r+\frac{\pi}{2} \right)\,,
\end{eqnarray}
Comparing this result with the boundary condition of the
scattering state wave function as $u(r \rightarrow \infty)
\rightarrow
2sin\left(kr-\frac{\pi}{2}\,\ell+\delta_{\ell}\right)$, we obtain
the phase shifts as
\begin{eqnarray}
\delta_{\ell}&=&\,\frac{\pi}{2}\,(1+\ell)+arg \Gamma(2i\kappa)-arg
\Gamma\left(\mu+i\kappa-
\sqrt{\frac{2m}{\lambda\hbar^2}\,(\epsilon-b)\,}\,\right)\nonumber\\&-&
arg \Gamma\left(\mu+i\kappa+
\sqrt{\frac{2m}{\lambda\hbar^2}\,(\epsilon-b)\,}\,\right)\,.
\end{eqnarray}
It is seen that the phase shifts of the Hellmann potential can be
produced by using the behavior of the hypergeometric functions at
infinity and they are dependent on the energy of the particle.

\section{Conclusion}
We have solved the Schrödinger equation for PT-/non-PT-symmetric
and non-Hermitian Hellmann potential for any angular momentum. The
normalized wave functions are obtained in terms of the
hypergeometric functions by using an approximation instead of the
centrifugal term. We have calculated energy eigenvalue. Its
numerical values for the bound states are listed in Table I. They
are compared with those of the previous results. We have seen that
our results are in good agreement especially for smaller parameter
values. The energy eigenvalue for the Coulomb potential is
obtained by setting potential parameters. Finally, we have studied
the scattering state solutions of the Hellmann potential and
obtained the phase shifts in terms of the angular momentum quantum
number $\ell$.

\section{Acknowledgments}
This research was partially supported by the Scientific and
Technical Research Council of Turkey.


\newpage

\begin{table}[htp]
\caption{The energy eigenvalues of Hellmann potential.}
\begin{tabular}{@{}ccllcccc@{}}
&&\multicolumn{3}{c}{$a=1\,\,b=0.5\,\,\lambda=0.001$}
&\multicolumn{3}{c}{$a=1\,\,b=-0.5\,\,\lambda=0.001$}
\\ \cline{3-5} \cline{6-8}
$n$ & $\ell$ & present & Ref. [11] & Ref. [28] & present & Ref. [11] & Ref. [28]   \\
1 & 0 & -0.25150 & -0.25100 & -0.25100 & -2.25050 & -2.24900 & -2.24900 \\
2 & 0 & -0.06400 & -0.06349 & -0.06349 & -0.56300 & -0.56150 & -0.56150 \\
  & 1 & -0.06375 & -0.06350 & -0.06350 & -0.56225 & -0.56150 & -0.56150 \\
3 & 0 & -0.02928 & -0.02876 & -0.02876 & -0.25050 & -0.24900 & -0.24900 \\
  & 1 & -0.02917 & -0.02877 & -0.02877 & -0.25017 & -0.24900 & -0.24900 \\
  & 2 & -0.02895 & -0.02877 & -0.02877 & -0.24950 & -0.24900 & -0.24900 \\
4 & 0 & -0.01713 & -0.01660 & -0.01660 & -0.14113 & -0.13963 & -0.13963 \\
  & 1 & -0.01706 & -0.01660 & -0.01660 & -0.14094 & -0.13963 & -0.13963 \\
  & 2 & -0.01694 & -0.01660 & -0.01660 & -0.14056 & -0.13963 & -0.13963 \\
  & 3 & -0.01675 & -0.01661 & -0.01660 & -0.14000 & -0.13963 & -0.13963 \\
&&\multicolumn{3}{c}{$a=1\,\,b=0.5\,\,\lambda=0.01$}
&\multicolumn{3}{c}{$a=1\,\,b=-0.5\,\,\lambda=0.01$}
\\ \cline{3-5} \cline{6-8}
$n$ & $\ell$ & present & Ref. [11] & Ref. [28] & present & Ref. [11] & Ref. [28]   \\
1 & 0 & -0.26502 & -0.25985 & -0.25985 & -2.25503 & -2.24000 & -2.24005\\
2 & 0 & -0.07760 & -0.07193 & -0.07193 & -0.56760 & -0.55270 & -0.55270\\
  & 1 & -0.07502 & -0.07197 & -0.07202 & -0.56002 & -0.55268 & -0.55266\\
3 & 0 & -0.04300 & -0.03657 & -0.03657 & -0.25522 & -0.24040 & -0.24044\\
  & 1 & -0.04180 & -0.03661 & -0.03664 & -0.25180 & -0.24042 & -0.24040\\
  & 2 & -0.03947 & -0.03665 & -0.03681 & -0.24502 & -0.24040 & -0.24034\\
4 & 0 & -0.03102 & -0.02367 & -0.02364 & -0.14602 & -0.13138 & -0.13138\\
  & 1 & -0.03031 & -0.02371 & -0.02371 & -0.14406 & -0.13137 & -0.13135\\
  & 2 & -0.02891 & -0.02374 & -0.02386 & -0.14016 & -0.13135 & -0.13129\\
  & 3 & -0.02690 & -0.02378 & -0.02404 & -0.13440 & -0.13134 & -0.13120\\
\end{tabular}
\end{table}

\newpage

\begin{figure}
\centering \subfloat[][Graphical representation of $1/x$
.]{\includegraphics[height=2.5in,
width=5.5in, angle=0]{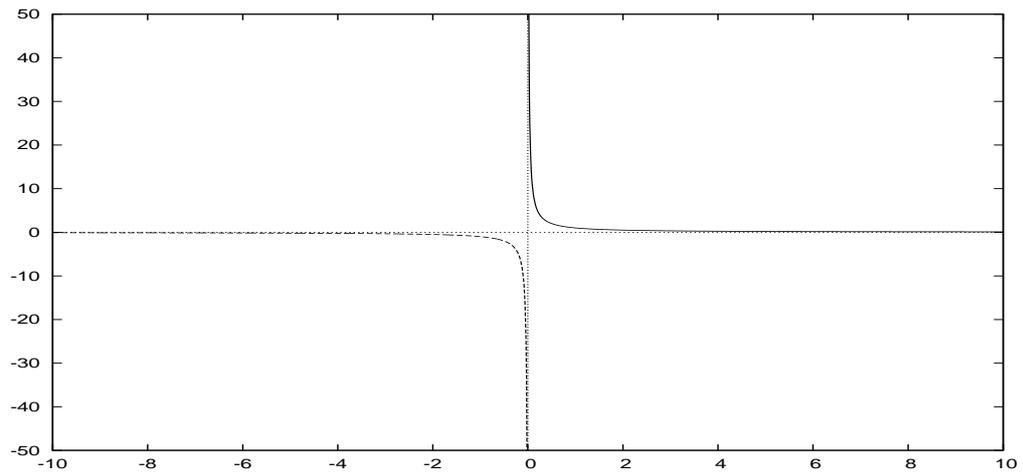}}\\
\subfloat[][Graphical representation of $\lambda/(1-e^{-\lambda
x})$ .]{\includegraphics[height=2.5in, width=5.5in,
angle=0]{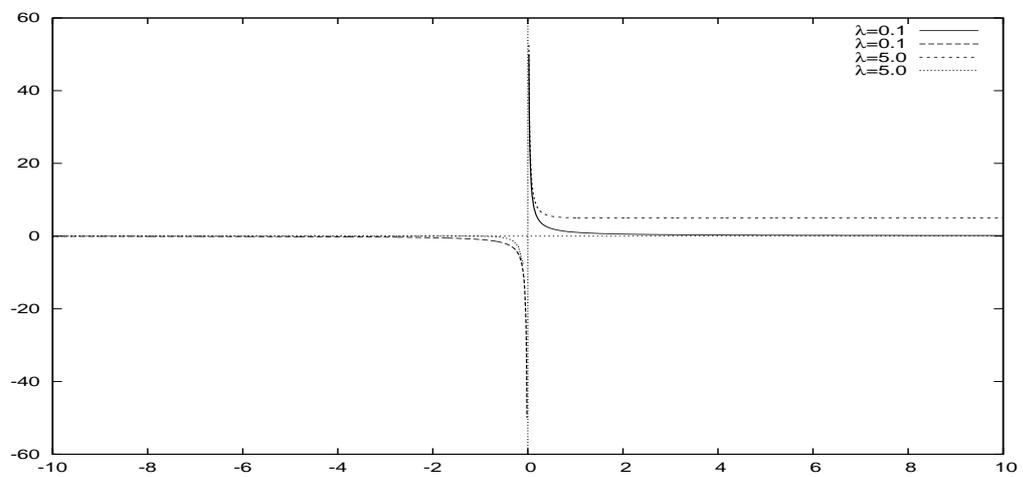}} \caption{Comparison of the functions $1/x$
and $\lambda/(1-e^{-\lambda x})$.}
\end{figure}

\end{document}